\documentclass{jfm}
\usepackage[T1]{fontenc}
\usepackage[latin9]{inputenc}
\usepackage{verbatim}
\usepackage{soul}
\usepackage{xcolor}
\usepackage{graphicx}
\usepackage{amssymb}
\usepackage{amsmath}
\usepackage{esint}
\usepackage[authoryear]{natbib}
\PassOptionsToPackage{normalem}{ulem}
\usepackage{ulem}

\makeatletter
\providecommand{\tabularnewline}{\\}
\providecolor{lyxadded}{rgb}{0,0,1}
\providecolor{lyxdeleted}{rgb}{1,0,0}

\def\vec#1{\mbox{\boldmath $\mathit{#1}$}}
\def\RE{\mathit{Re}}
\def\Rm{\mathit{Rm}}
\def\Ha{\mathit{Ha}}
\def\i{\mathrm{i}}

\makeatother

\begin{document}

\title[Linear stability of MHD flow in a square duct with thin conducting walls]
{Linear stability of magnetohydrodynamic flow in a square duct with thin conducting walls}

\author[J. Priede, T. Arlt, L. Bühler]{J\={a}nis Priede\aff{1}, Thomas Arlt\aff{2} \and~Leo Bühler\aff{2}}
\affiliation{
\aff{1}Applied Mathematics Research Centre,\\
Coventry University, UK\\
\aff{2}Institut für Kern- und Energietechnik,\\
Karlsruhe Institute of Technology, Germany}

\maketitle
\begin{abstract}
\global\long\def\Ha{\mathit{Ha}}
\global\long\def\RE{\mathit{Re}}
This study is concerned with the numerical linear stability analysis
of liquid-metal flow in a square duct with thin electrically conducting
walls subject to a uniform transverse magnetic field. We derive an
asymptotic solution for the base flow that is valid not only for high
but also moderate magnetic fields. This solution shows that for low
wall conductance ratios $c\ll1,$ an extremely strong magnetic field
with Hartmann number $\Ha\sim c^{-4}$ is required to attain the asymptotic
flow regime considered in the previous studies. We use a vector stream
function--vorticity formulation and a Chebyshev collocation method
to solve the eigenvalue problem for three-dimensional small-amplitude
perturbations in ducts with realistic wall conductance ratios $c=1,$
and $0.1$ and $0.01$ and Hartmann numbers up to $10^{4}.$ As for
similar flows, instability in a sufficiently strong magnetic field
is found to occur in the sidewall jets with the characteristic thickness
$\delta\sim\Ha^{-1/2}.$ This results in the critical Reynolds number
and wavenumber increasing asymptotically with the magnetic field as
$\RE_{c}\sim110\Ha^{1/2}$ and $k_{c}\sim0.5\Ha^{1/2}.$ The respective
critical Reynolds number based on the total volume flux in a square
duct with $c\ll1$ is $\bar{\RE}_{c}\approx520.$ Although this value
is somewhat larger than$\bar{\RE}_{c}\approx313$ found by \citet{Ting1991}
for the asymptotic sidewall jet profile, it still appears significantly
lower than the Reynolds numbers at which turbulence is observed in
experiments as well as in direct numerical simulations of this type
of flow.
\end{abstract}

\section{Introduction}

Magnetohydrodynamic (MHD) flows in ducts, which have been studied
for almost 80 years \citep{Hartmann1937,Hartmann1937a}, are still
a subject of significant scientific interest \citep{Krasnov2013,Zikanov2014}.
This is mainly due to their potential application in the liquid-metal
cooling blankets of prospective fusion reactors \citep{Buehler2007}.
Of particular interest are the stability and transition to turbulence
in these flows, may strongly affect their transport properties. Turbulent
mixing can enhance not only the transport of heat and mass but also
that of momentum, which, in turn, would lead to an increased hydrodynamic
resistance of the duct. 

The magnetic field can have a diverse effect on the flow of a conducting
liquid. Usually, the magnetic field suppresses and stabilizes the
flow by ohmic dissipation resulting from the induced electric current,
as in the classical case of a conducting liquid heated from below
\citep{Chandrasekhar1961}. But ohmic dissipation can also destabilize
some rotational flows through the so-called helical magneto-rotational
instability mechanism, which does not directly affect the base flow
\citep{Hollerbach2005,Priede2007}. There are yet there are other
flows that can be destabilized directly by the magnetic field modifying
their velocity profiles, as in ducts with conducting walls, where
highly unstable sidewall jets are often created \citep{Chang1961,Uflyand1961,Hunt1965}.
The flow in a duct with thin conducting walls, which is at the focus
of this study, belongs to the latter type \citep{Walker1981}.

The linear stability of two such flows have been comprehensively analysed
previously and found to be highly unstable. The first flow was in
a duct with perfectly conducting walls perpendicular to the magnetic
field and insulating parallel walls, i.e. so-called Hunt's flow \citep{Priede2010},
whereas the second was in a duct with all walls perfectly conducting
\citep{Priede2012}. The low stability of these flows is due to the
velocity jets that develop along the walls parallel to the magnetic
field when the latter is strong enough. In the duct with perfectly
conducting walls, the jets are relatively weak, with maximal velocity
only slightly higher than that of the core flow. In Hunt's flow, on
the contrary, the velocity jets are particularly strong, with maximal
velocity increasing relative to that of the core flow directly with
the Hartmann number $\Ha$ \citep{Hunt1965}.

The perfectly conducting or insulating walls assumed in the previous
two studies are rather far from reality, where the walls usually have
a finite electrical conductivity. Most common are ducts with thin
conducting walls. Jets as in Hunt's flow, but with a lower relative
velocity $\sim\Ha^{1/2},$ can form also in these ducts \citep{Walker1981}.
In contrast to Hunt's flow, where the jets carry the dominant part
of the volume flux, in ducts with thin conducting walls they carry
only a fraction of the total volume flux. In a sufficiently strong
magnetic field, the total volume flux depends solely on the ratio
of the wall conductance to that of the liquid layer, which is subsequently
referred to as the wall conductance ratio. In this study, we show,
that for low wall conductance ratios $c\ll1,$ an extremely strong
magnetic field with $\Ha\sim c^{-4}$ is required to attain this asymptotic
flow regime. The linear stability of such an asymptotic sidewall jet
has been studied by \citet{Ting1991}. In this approximation, the
electrical resistance of the walls is assumed to be much higher than
that of the liquid metal in the duct but much lower than that of the
MHD boundary layers that form along the walls in a strong magnetic
field. The aim of the present study is to investigate the linear stability
of this flow for realistic wall conductance ratios in a magnetic field
of feasible strength. 

The paper is organized as follows. The problem is formulated in $\S$\ref{sec:prob}
and the numerical method is outlined in \ref{sec:num} In $\S$\ref{sec:res}
we present and discuss numerical results for a square duct in a vertical
magnetic field. The paper is concluded with a discussion of results
in $\S$\ref{sec:sum}. An asymptotic solution for the base flow valid
in a wide range of magnetic field strength is presented in the appendix
\ref{sec:bflow}.

\begin{figure}
\begin{centering}
\includegraphics[bb=100bp 80bp 400bp 280bp,clip,width=0.6\columnwidth]{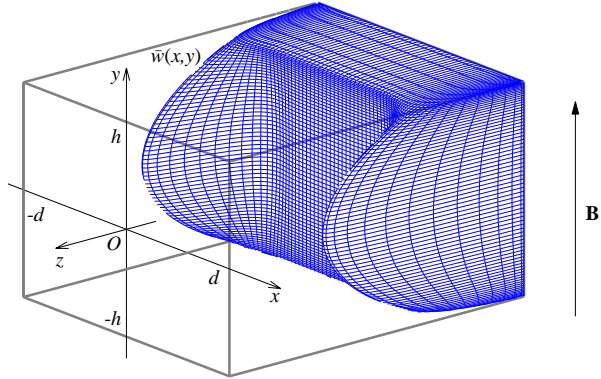} 
\par\end{centering}

\caption{\label{fig:sketch}The base flow profile in a square duct with thin
conducting walls subject to a vertical magnetic field for $c=0.1$
and $\protect\Ha=100.$ }
\end{figure}

\section{\label{sec:prob}Formulation of the problem}

Consider the flow of an incompressible viscous electrically conducting
liquid with density $\rho,$ kinematic viscosity $\nu$ and electrical
conductivity $\sigma$ driven by a constant gradient of pressure $p$
applied along the axis of straight duct of rectangular cross-section
with half-width $d$ and half-height $h$ subject to a transverse
homogeneous magnetic field $\vec{B}.$ The walls of the duct are assumed
to have small thickness $d_{w}\ll d$ and electrical conductivity
$\sigma_{w}.$ 

The liquid flow is governed by the Navier-Stokes equation
\begin{equation}
\partial_{t}\vec{v}+(\vec{v}\cdot\vec{\nabla})\vec{v}=-\rho^{-1}\vec{\nabla}p+\nu\vec{\nabla}^{2}\vec{v}+\rho^{-1}\vec{f},\label{eq:NS}
\end{equation}
with the electromagnetic body force $\vec{f}=\vec{j}\times\vec{B}$
involving the induced electric current $\vec{j},$ which is governed
by Ohm's law for a moving medium, 
\begin{equation}
\vec{j}=\sigma(\vec{E}+\vec{v}\times\vec{B}).\label{eq:Ohm}
\end{equation}
 The flow is assumed to be sufficiently slow for the induced magnetic
field to be negligible relative to the imposed one. This corresponds
to the so-called inductionless approximation, which holds for small
magnetic Reynolds numbers \global\long\def\Rm{\mathit{Rm}}
$\Rm=\mu_{0}\sigma v_{0}d\ll1,$ where $\mu_{0}$ is the permeability
of free space and $v_{0}$ is a characteristic velocity of the flow.
In addition, we assume the characteristic time of velocity variation
to be much longer than the magnetic diffusion time $\tau_{m}=\mu_{0}\sigma d^{2}.$
This is known in MHD as the quasi-stationary approximation \citep{Roberts1967},
which leads to $\vec{E}=-\vec{\nabla}\phi,$ where $\phi$ is the
electrostatic potential. The velocity and current satisfy mass and
charge conservation, $\vec{\nabla}\cdot\vec{v}=0,\vec{\nabla}\cdot\vec{j}=0.$
Applying the latter to Ohm's law (\ref{eq:Ohm}) and using the inductionless
approximation, we obtain 
\begin{equation}
\vec{\nabla}^{2}\phi=\vec{B}\cdot\vec{\omega},\label{eq:phi}
\end{equation}
where $\vec{\omega}=\vec{\nabla}\times\vec{v}$ is the vorticity.
At the duct walls $S$, the normal $(n)$ and tangential $(\tau)$
velocity components satisfy the impermeability and no-slip boundary
conditions $\left.v_{n}\right\vert _{s}=0$ and $\left.v_{\tau}\right\vert _{s}=0.$
Charge conservation applied to the thin wall leads to the following
boundary condition, $\left.\partial_{n}\phi-dc\vec{\nabla}_{\tau}^{2}\phi\right\vert _{s}=0,$
where $c=\sigma_{w}d_{w}/(\sigma d)$ is the wall conductance ratio
\citep{Walker1981}. At the corner point $P$, this condition reduces
to that of the continuity of the tangential current in the wall, i.e.
$\left[dc\partial_{\tau}\phi\right]_{P}=0,$ where the square brackets
denote the jump in the enclosed quantity. 

We employ Cartesian coordinates, with the origin set at the centre
of the duct, $x$-, $y$- and $z$-axes directed along its width,
height and length, respectively, as shown in figure \ref{fig:sketch},
and the velocity defined as $\vec{v}=(u,v,w).$ The problem admits
a purely rectilinear base flow with a single velocity component along
the duct $\bar{\vec{v}}=(0,0,\bar{w}(x,y)),$ which is shown in figure
\ref{fig:sketch} for $c=0.1$ and $\Ha=100.$

In the following, all variables are non-dimensionalized by using the
maximum velocity $\bar{w}_{0}$ and the half-width of the duct $d$
as the velocity and length scales. The time, pressure, magnetic field
and electrostatic potential are scaled by $d^{2}/\nu,$ $\rho\bar{w}_{0}^{2},$
$B=\left\vert \vec{B}\right\vert $ and $\bar{w}_{0}dB,$ respectively.
Note that we use the maximum rather than average velocity of the base
flow as the characteristic scale because the stability of this flow
is determined by the former. 

The base flow can conveniently be determined using the $z$-component
of the induced magnetic field $\bar{b}$ instead of the electrostatic
potential $\bar{\phi}.$ Then the governing equations for the base
flow take the form 
\begin{eqnarray}
\vec{\nabla}^{2}\bar{w}+\Ha\partial_{y}\bar{b} & = & \bar{P,}\label{eq:wbar}\\
\vec{\nabla}^{2}\bar{b}+\Ha\partial_{y}\bar{w} & = & 0,\label{eq:bbar}
\end{eqnarray}
where $\Ha=dB\sqrt{\sigma/(\rho\nu)}$ is the Hartmann number and
$\bar{b}$ is scaled by $\mu_{0}\sqrt{\sigma\rho\nu^{3}}/d.$ The
constant dimensionless axial pressure gradient $\bar{P}$ that drives
the flow is determined from the normalization condition $\bar{w}_{\max}=1.$
The velocity satisfies the no-slip boundary condition $\bar{w}=0$
at $x=\pm1$ and $y=\pm A,$ where $A=h/d$ is the aspect ratio, which
is set equal to $1$ for the square cross-section duct considered
in this study. The boundary condition for the induced magnetic field
at the thin wall is 
\begin{equation}
\bar{b}=c\partial_{n}\bar{b}.\label{bc:bbar}
\end{equation}

The linear stability of the flow is analysed using a non-standard
vector streamfunction-vorticity formulation \citep{Priede2010,Priede2012}.
Following this approach, a vector streamfunction $\vec{\psi}$ is
introduced to satisfy the incompressibility constraint $\vec{\nabla}\cdot\vec{v}=0$
for the flow perturbation by seeking the velocity distribution in
the form $\vec{v}=\vec{\nabla}\times\vec{\psi}.$ Each component of
$\vec{\psi}$ describes a solenoidal flow with two velocity components
in the plane transverse to that component where the isolines of the
respective component of $\vec{\psi}$ represent the streamlines of
that flow. Since $\vec{\psi}$ is determined up to a gradient of arbitrary
function, an additional constraint 
\begin{equation}
\vec{\nabla}\cdot\vec{\psi}=0,\label{eq:divpsi}
\end{equation}
which is analogous to the Coulomb gauge for the magnetic vector potential
$\vec{A}$ \citep{Jackson1998}, can be applied. Similar to the incompressibility
constraint for $\vec{v},$ this gauge leaves only two independent
components of $\vec{\psi}.$ 

The pressure gradient is eliminated by taking the \textit{curl} of
the dimensionless counterpart of (\ref{eq:NS}), which leads to the
following equations for $\vec{\psi}$ and $\vec{\omega}$: 
\begin{eqnarray}
\partial_{t}\vec{\omega} & = & \vec{\nabla}^{2}\vec{\omega}-\RE\vec{g}+\Ha^{2}\vec{h},\label{eq:omeg}\\
0 & = & \vec{\nabla}^{2}\vec{\psi}+\vec{\omega},\label{eq:psi}
\end{eqnarray}
where $\RE=\bar{w}_{0}d/\nu$ is the Reynolds number based on the
maximum jet velocity $\bar{w}_{0}$,and $\vec{g}=\vec{\nabla}\times(\vec{v}\cdot\vec{\nabla})\vec{v},$
and $\vec{h}=\vec{\nabla}\times\vec{f}$ are the \textit{curls} of
the dimensionless convective inertial and electromagnetic forces,
respectively. 

The boundary conditions for $\vec{\psi}$ and $\vec{\omega}$ are
derived as follows. The impermeability condition applied integrally
as $\int_{s}\vec{v}\cdot\vec{\mathrm{d}s}=\oint_{l}\vec{\psi}\cdot\vec{\mathrm{d}l}=0$
to an arbitrary area of wall $s$ encircled by a contour $l$ yields
$\left.\psi_{\tau}\right|_{s}=0.$ This boundary condition substituted
into (\ref{eq:divpsi}) results in $\left.\partial_{n}\psi_{n}\right|_{s}=0.$
In addition, the no-slip condition applied integrally, $\oint_{l}\vec{v}\cdot\vec{\mathrm{d}l}=\int_{s}\vec{\omega}\cdot\vec{\mathrm{d}s}$,
yields $\left.\omega_{n}\right|_{s}=0.$

The linear stability of the base flow $\{\bar{\vec{\psi}},\bar{\vec{\omega}},\bar{\phi}\}(x,y)$
is analysed with respect to infinitesimal disturbances in the standard
form of harmonic waves travelling along the axis of the duct, \global\long\def\i{\mathrm{i}}
 
\[
\{\vec{\psi},\vec{\omega},\phi\}(\vec{r},t)=\{\bar{\vec{\psi}},\bar{\vec{\omega}},\bar{\phi}\}(x,y)+\{\hat{\vec{\psi}},\hat{\vec{\omega}},\hat{\phi}\}(x,y)e^{\lambda t+\i kz},
\]
 where $k$ is a real wavenumber and $\lambda$ is, in general, a
complex growth rate. This expression substituted into (\ref{eq:omeg},\ref{eq:psi})
results in 
\begin{eqnarray}
\lambda\hat{\vec{\omega}} & = & \vec{\nabla}_{k}^{2}\hat{\vec{\omega}}-\RE\hat{\vec{g}}+\Ha^{2}\hat{\vec{h}},\label{eq:omegh}\\
0 & = & \vec{\nabla}_{k}^{2}\hat{\vec{\psi}}+\hat{\vec{\omega}},\label{eq:psih}\\
0 & = & \vec{\nabla}_{k}^{2}\hat{\phi}-\hat{\omega}_{\shortparallel},\label{eq:phih}
\end{eqnarray}
 where  and$\shortparallel$ and $\perp$ respectively denote the
components along and transverse to the magnetic field in the $(x,y)$-plane.
Because of the solenoidality of $\hat{\vec{\omega}},$ it suffices
to consider only the $x$- and $y$-components of (\ref{eq:omegh}),
which contain $\hat{h}_{\perp}=-\partial_{xy}\hat{\phi}-\partial_{\shortparallel}\hat{w},$
$\hat{h}_{\shortparallel}=-\partial_{\shortparallel}^{2}\hat{\phi}$
and 
\begin{eqnarray}
\hat{g}_{x} & = & k^{2}\hat{v}\bar{w}+\partial_{yy}(\hat{v}\bar{w})+\partial_{xy}(\hat{u}\bar{w})+\i2k\partial_{y}(\hat{w}\bar{w}),\label{eq:gx}\\
\hat{g}_{y} & = & -k^{2}\hat{u}\bar{w}-\partial_{xx}(\hat{u}\bar{w})-\partial_{xy}(\hat{v}\bar{w})-\i2k\partial_{x}(\hat{w}\bar{w}),\label{eq:gy}
\end{eqnarray}
 where 
\begin{eqnarray}
\hat{u} & = & \i k^{-1}(\partial_{yy}\hat{\psi}_{y}-k^{2}\hat{\psi}_{y}+\partial_{xy}\hat{\psi}_{x}),\label{eq:uh}\\
\hat{v} & = & -\i k^{-1}(\partial_{xx}\hat{\psi}_{x}-k^{2}\hat{\psi}_{x}+\partial_{xy}\hat{\psi}_{y}),\label{eq:vh}\\
\hat{w} & = & \partial_{x}\hat{\psi}_{y}-\partial_{y}\hat{\psi}_{x}.\label{eq:wh}
\end{eqnarray}
 The relevant boundary conditions are
\begin{eqnarray}
\partial_{x}^{2}\hat{\phi}-k^{2}\hat{\phi}-c^{-1}\partial_{y}\hat{\phi}=\hat{\psi}_{y}=\partial_{x}\hat{\psi}_{x}=\partial_{x}\hat{\psi}_{y}-\partial_{y}\hat{\psi}_{x}=\hat{\omega}_{x}=0 & \mbox{at} & x=\pm1,\label{eq:bcx}\\
\partial_{y}^{2}\hat{\phi}-k^{2}\hat{\phi}-c^{-1}\partial_{x}\hat{\phi}=\hat{\psi}_{x}=\partial_{y}\hat{\psi}_{y}=\partial_{x}\hat{\psi}_{y}-\partial_{y}\hat{\psi}_{x}=\hat{\omega}_{y}=0 & \mbox{at} & y=\pm A.\label{eq:bcy}
\end{eqnarray}

\section{\label{sec:num}Numerical procedure}

The problem was solved by a spectral collocation method on a Chebyshev-Lobatto
grid with even number of points $2N_{x}+2$ and $2N_{y}+2$ in the
$x$- and $y$-directions, where $N_{x,y}=35\cdots70$ were used for
various combinations of the control parameters to achieve accuracy
of at least three significant figures. Accuracy was verified by recalculating
dubious results with a higher numerical resolution using typically
five more collocation points in each direction. The number of collocation
points required for a sufficiently accurate solution was found to
increase with the Hartmann number reaching $N_{x,y}=70$ at $\Ha=10^{4}.$
Owing to the double reflection symmetry of the base flow with respect
to the $x=0$ and $y=0$ planes, small-amplitude perturbations with
different parities in $x$ and $y$ decouple from each other. This
results in four mutually independent modes, which we classify as $(o,o),$
$(o,e),$  and $(e,e)$ according to whether the $x$ and $y$ symmetry
of $\hat{\psi}_{x}$ is odd or even, respectively. Our classification
of modes corresponds to the symmetries I, II, III and IV used by \citet{Tatsumi1990}
and \citet{Uhlmann2006} (see table \ref{tab:mod}). The symmetry
allows us to solve the linear stability problem for each of the four
modes separately using only one quadrant of the duct cross-section
with $N_{x}\times N_{y}$ internal collocation points.

The four modes listed in table \ref{tab:mod} have the following spatial
structure. Mode I has a $y$-component of streamfunction-vorticity
that which is even in both the $x$- and $y-$directions, i.e. mirror-symmetric
with respect to both mid-planes. This corresponds to vertical vortices
that rotate in the same sense in all four quadrants of the duct. Such
vortices can span the height of the duct and thus, can be relatively
uniform along a vertical magnetic field. It also means that the vortices
at the opposite sidewalls rotate in the same sense and thus can enhance
each other through a shared flux across the vertical mid-plane of
the duct. Mode II differs from mode I by the opposite vertical symmetry.
It means that the vertical vortices associated with this mode change
their direction of rotation at the horizontal mid-plane and rotate
in the opposite senses in the top and bottom parts of the duct. Such
vortices are inherently non-uniform along the height of the duct and
thus subject to a strong damping by the vertical magnetic field. Modes
III and IV differ from modes I and II by the opposite spanwise symmetry.
It means that the vertical vortices at the opposite sidewalls for
these two modes rotate in the opposite senses and are separated from
each other by the vertical mid-plane of the duct. The symmetry of
the $x$-component of streamfunction-vorticity, which is associated
with spanwise vortices, is always opposite to that of the $y$-component.
This follows from the symmetry of the associated velocity components,
which are defined in terms of $\hat{\psi}_{x}$ and $\hat{\psi}_{y}$
by (\ref{eq:uh}-\ref{eq:wh}). Note that modes I and IV differ from
one another only by a $90^{\circ}$ rotation around the $z$-axis,
which is not the case for modes II and III.

\begin{table}
\begin{centering}
\begin{tabular}{rcccc}
 & I & II & III & IV\tabularnewline
$\hat{\psi}_{x},\,\hat{\omega}_{x},\,\hat{v}:$ & $(o,o)$ & $(o,e)$ & $(e,o)$ & $(e,e)$\tabularnewline
$\hat{w}:$ & $(o,e)$ & $(o,o)$ & $(e,e)$ & $(e,o)$\tabularnewline
$\hat{\psi}_{z},\,\hat{\omega}_{z}:$ & $(e,o)$ & $(e,e)$ & $(o,o)$ & $(o,e)$\tabularnewline
$\hat{\psi}_{y},\,\hat{\omega}_{y},\,\hat{u},\,\phi:$ & $(e,e)$ & $(e,o)$ & $(o,e)$ & $(o,o)$\tabularnewline
\end{tabular}
\par\end{centering}

\caption{\label{tab:mod}The $(x,y)$ parities of different variables for symmetries
I, II, III and IV; $e$ - even, $o$ - odd }
\end{table}

The numerical procedure employed in this study is basically the same
as that used by \citet{Priede2010}, where a detailed description
and validation of the code are presented. The only major difference
is that the original code was designed for perfectly electrically
conducting and insulating walls, whereas here walls with a finite
electrical conductivity are considered. This leads to different boundary
conditions for both the induced magnetic field (\ref{bc:bbar}) and
the electric potential (\ref{eq:bcx}), (\ref{eq:bcy}). Owing to
the fact that the Laplace operator on the l.h.s of (\ref{eq:phih})
turns into the electric boundary condition when the second-order partial
derivative normal to the boundary is substituted by the respective
first-order derivative divided by negative wall conductivity ratio
as in (\ref{eq:bcx}) and (\ref{eq:bcy}), the electric potential
can efficiently be eliminated from the problem by using the matrix
diagonalization algorithm \citep[Sec. 4.1.4]{Canuto2007} to invert
the discretized counterpart of (\ref{eq:phih}).

\section{\label{sec:res}Results}

\subsection{Base flow}

\begin{figure}
\begin{centering}
\includegraphics[bb=115bp 90bp 370bp 260bp,clip,width=0.51\textwidth]{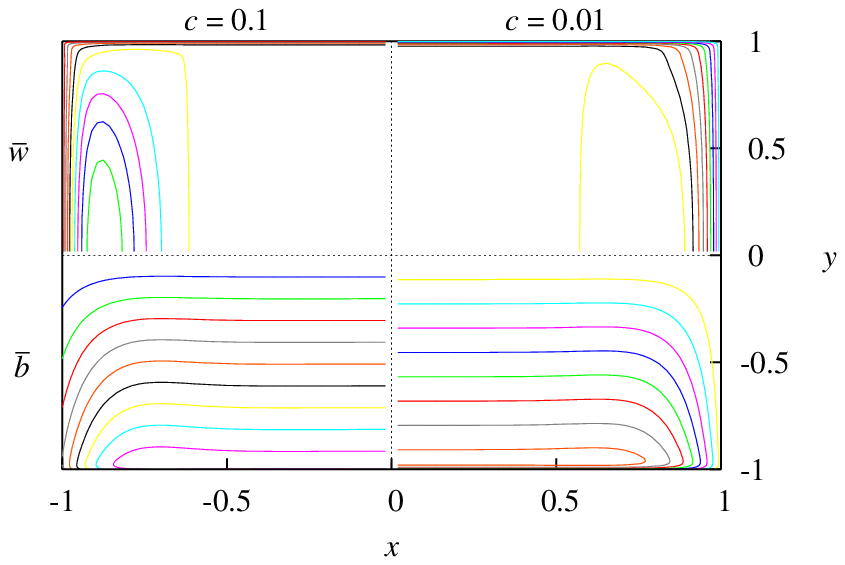}\put(-190,130){(\textit{a})}\includegraphics[width=0.49\textwidth]{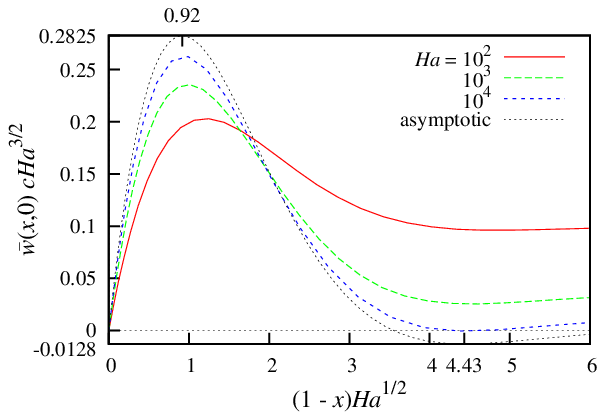}\put(-160,130){(\textit{b})} 
\par\end{centering}

\caption{\label{fig:bflow} (\emph{a}) Isolines of the base flow velocity $(y>0)$
and the electric current lines $(y<0)$ for $c=0.1$ $(x<0)$ and
$c=0.01$ $(x>0)$ at $\protect\Ha=100$ shown in the respective quadrants
of the duct cross-section; and (\emph{b})the horizontal base flow
velocity profiles in the vicinity of the side-wall at $y=0$ in stretched
coordinate $(1-x)\protect\Ha^{1/2}$ for various Hartmann numbers
and $c=0.1.$ }
\end{figure}

\begin{figure}
\begin{centering}
\includegraphics[width=0.5\textwidth]{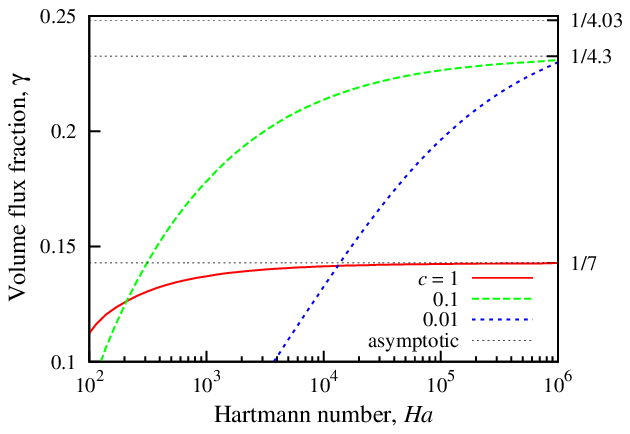}\put(-165,135){(\textit{a})}\includegraphics[width=0.5\textwidth]{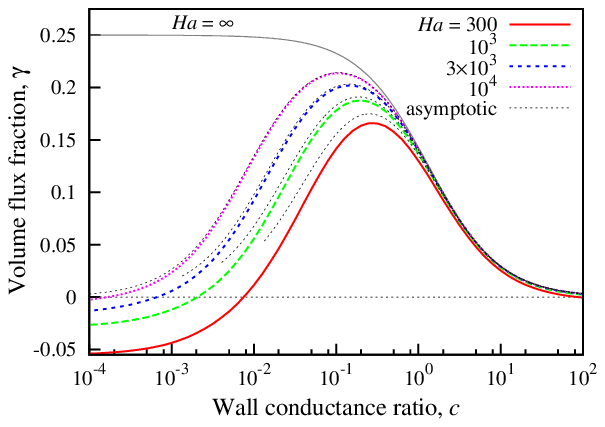}\put(-165,135){(\textit{b})} 
\par\end{centering}

\begin{centering}
\includegraphics[width=0.5\textwidth]{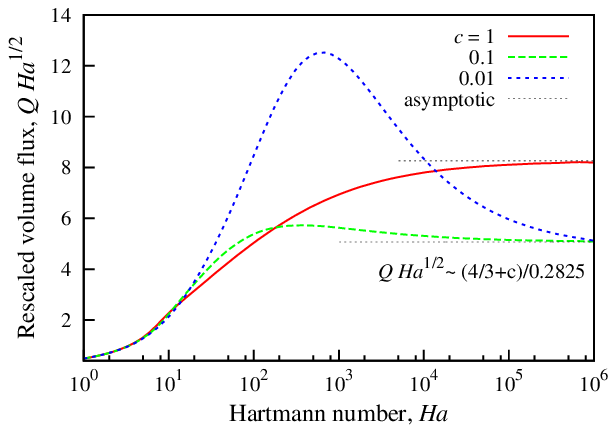}\put(-165,135){(\textit{c})}
\par\end{centering}

\caption{\label{fig:flw} The volume flux fraction $\gamma$ carried by the
side-wall jets (\emph{a}) depending on the Hartmann number for fixed
wall conductance ratios $c=1,0.1,0.01,$ and (\emph{b}) versus the
wall conductance ratio for various Hartmann numbers with the asymptotic
solution (\ref{eq:gam}) shown by the thin dotted lines; (\emph{c})
rescaled total volume flux $Q\protect\Ha^{1/2}$ versus the Hartmann
number for the base flow normalized with maximal jet velocity with
the asymptotic solution (\ref{eq:Q}) shown by the thin dotted lines. }
\end{figure}

Let us first consider the characteristics of the base flow that are
pertinent to its stability. The most prominent feature of the base
flow is the high-velocity jets that form in the strong magnetic field
along the sidewalls of the duct provided that the conductance ratio
is sufficiently high for the given magnetic field strength. This effect
is illustrated by figures \ref{fig:sketch} and \ref{fig:bflow}(a),
where the jets are seen to be pronounced at $\Ha=100$ and $c=0.1$
and then to become relatively weak at the wall conductance ratio $c=0.01.$
The velocity of the jets relative to that of the core flow typically
increases as $\sim\Ha^{1/2}$ while their thickness reduces as $\sim\Ha^{-1/2}.$
As seen in figure \ref{fig:bflow}(b), the velocity maximum 
\begin{equation}
\bar{w}_{\max}\sim-0.2825c^{-1}\Ha^{-3/2}\bar{P},\label{eq:wmax}
\end{equation}
which follows from the asymptotic solution (\ref{eq:w0}) presented
in the appendix \ref{sec:bflow}, is located at distance $\delta\sim0.92\Ha^{-1/2}$
from the sidewall. The velocity in the core of the flow defined by
(\ref{eq:winf}) is $\bar{w}_{\textrm{core}}\sim-(1+c^{-1})\Ha^{-2}\bar{P}.$
Thus, the fraction of the volume flux carried by the jets is comparable
to that carried by the core flow. In a strong magnetic field, this
fraction is expected to approach a constant determined solely by $c.$
As seen in figure \ref{fig:flw}(a), an extremely strong magnetic
field is required to achieve this asymptotic flow regime. For example,
the volume flux fraction carried by the the sidewall jets for $c=0.01$
at $\Ha=10^{6}$ is still approximately $8\%$ below its asymptotic
value $\gamma=\frac{1}{4.03},$ which follows from (\ref{eq:ginf}).
For a fixed $\Ha,$ $\gamma$ is seen in figure \ref{fig:flw}(b)
to attain a maximum at $c\approx\Ha^{-1/4}$ in a good agreement with
the asymptotic solution (\ref{eq:cmax}). It is important to note
that, at this point, $\gamma$ starts to deviate from its asymptotic
value for $\textit{Ha}\rightarrow\infty$ when $c$ is reduced. Thus,
for $\gamma$ to reach its high-field asymptotic limit (\ref{eq:ginfw})
at $c\ll1,$ an extremely strong magnetic field with $\Ha\sim c^{-4}$
is required. For the base flow normalized with the maximal velocity,
which is used as the characteristic velocity in this study, the total
volume flux is seen in figure \ref{fig:flw}(c) to approach the value
predicted by the asymptotic solution, 
\begin{equation}
Q\sim\left(\tfrac{4}{3}+c\right)/(0.2825\Ha^{1/2}),\label{eq:Q}
\end{equation}
where the normalization coefficient at $\Ha^{1/2}$ follows from (\ref{eq:wmax}).
This relation can be used to rescale our results with the average
velocity in a sufficiently strong magnetic field. We use an overbar
to distinguish the Reynolds number based on the total volume flux
from that introduced earlier using the maximal jet velocity.

\subsection{Linear stability}

\begin{figure}
\begin{centering}
\includegraphics[width=0.5\textwidth]{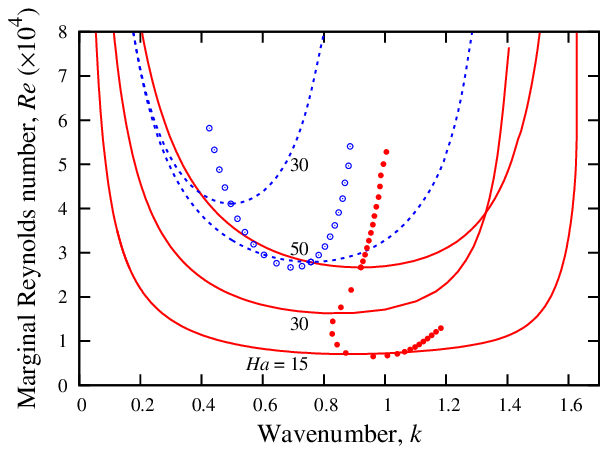}\put(-165,135){(\textit{a})}\includegraphics[width=0.5\textwidth]{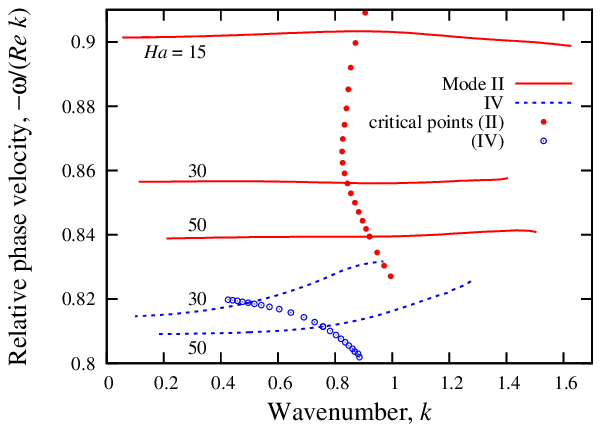}\put(-165,135){(\textit{b})} 
\par\end{centering}

\caption{\label{fig:rewk_Ge1_12}The marginal Reynolds number (\emph{a}) and
the relative phase velocity (\emph{b}) versus the wavenumber for neutrally
stable modes of type II and IV in a square duct $(A=1)$ with the
wall conductance ratio $c=1$ at various Hartmann numbers which are
shown next to the curves. Dots indicate the critical points, which
correspond to the lowest marginal $\protect\RE$ for each $\protect\Ha;$
these points cover a broader range of $\protect\Ha$ than the marginal
stability curves.}
\end{figure}

\begin{figure}
\begin{centering}
\includegraphics[width=0.5\textwidth]{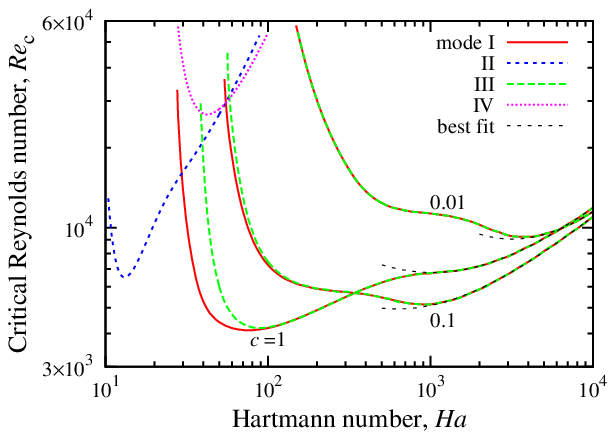}\put(-160,25){(\textit{a})}\includegraphics[width=0.5\textwidth]{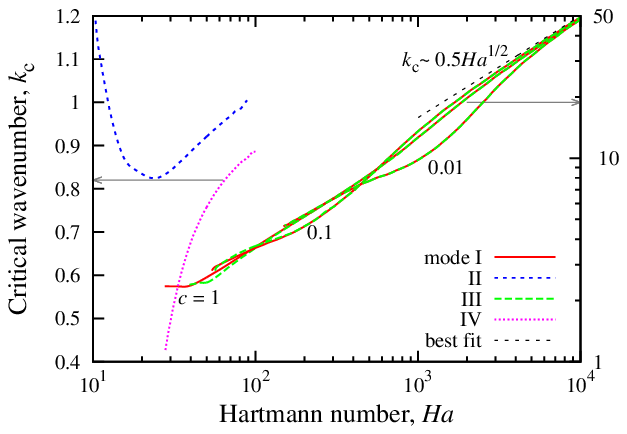}\put(-165,25){(\textit{b})}\\
\includegraphics[width=0.5\textwidth]{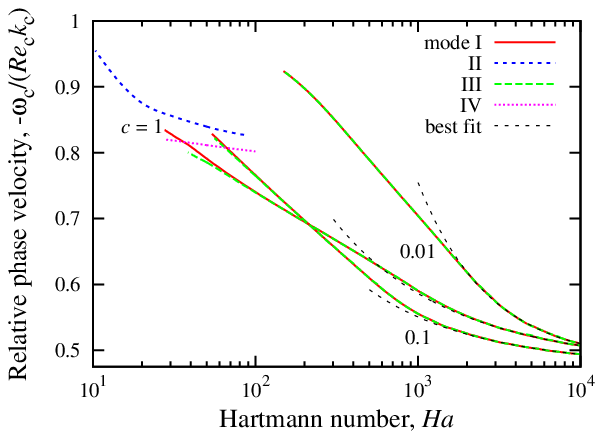}\put(-165,25){(\textit{c})} 
\par\end{centering}

\caption{\label{fig:rec_Ha_Ge1c}The critical Reynolds number (\emph{a}), wavenumber
(\emph{b}) and relative phase velocity (\emph{c}) against the Hartmann
number for wall conductance ratios $c=1,$ 0.1, 0.01. Modes II and
IV are shown only for $c=1.$ }
\end{figure}

The following results are for the flow in a square duct $(A=1)$,
which is \emph{linearly} stable without the magnetic field \citep{Tatsumi1990}.
We start with a moderate wall conductance ratio $c=1$ at which the
flow is expected to be similar to that in a perfectly conducting duct.
As in the perfectly conducting duct \citep{Priede2012}, the magnetic
field renders the flow linearly unstable at $\Ha\gtrsim10$ with respect
to a mode of symmetry type II (see table \ref{tab:mod}). The vorticity
component along the magnetic field of this mode is an odd function
in the field direction and an even function spanwise. The marginal
Reynolds number, at which the growth rate turns to zero $(\Re[\lambda]=0),$
and the associated relative phase speed $-\omega/(\RE k)$, which
is defined by the imaginary part of the growth rate $\omega=\Im[\lambda]$
relative to the maximum jet velocity $\RE$, are plotted against the
wavenumber in figure \ref{fig:rewk_Ge1_12}. The lowest point on each
marginal Reynolds number curve defines a critical Reynolds number
$\RE_{c}$ at which the flow first becomes unstable at the given Hartmann
number. These points are marked by dots in figure \ref{fig:rewk_Ge1_12}
and also plotted in figure \ref{fig:rec_Ha_Ge1c} against the Hartmann
number. In figure \ref{fig:rec_Ha_Ge1c}, the mode II instability
is seen to appear at $\Ha\approx10$ and very high $\RE_{c}.$ With
increase of the Hartmann number, $\RE_{c}$ quickly drops to the minimum
$\RE_{c}\approx6500$ at $\Ha\approx13$ and then starts to increase,
reaching $\RE_{c}\approx6\times10^{4}$ at $\Ha=100.$ The steep stabilization
of this mode is due to its antisymmetric vorticity distribution along
the magnetic field, which is smoothed out and thus efficiently suppressed
by the magnetic field.

\begin{figure}
\begin{centering}
\includegraphics[width=0.5\textwidth]{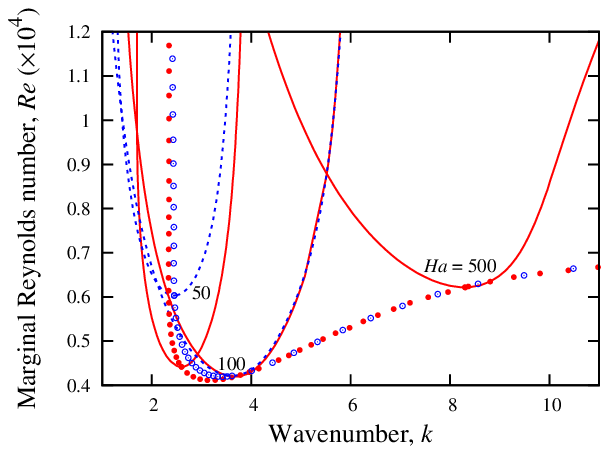}\put(-165,135){(\textit{a})}\includegraphics[width=0.5\textwidth]{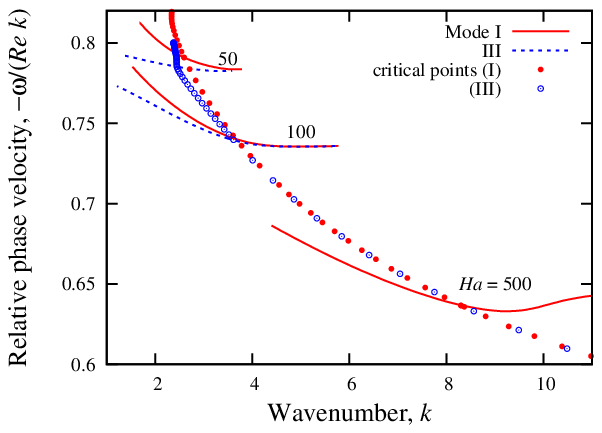}\put(-165,135){(\textit{b})}
\par\end{centering}

\caption{\label{fig:rewk_Ge1_11-21}The marginal Reynolds number (\emph{a})
and the relative phase velocity (\emph{b}) versus the wavenumber for
neutrally stable modes of type I and III in a square duct $(A=1)$
with the wall conductance ratio $c=1$ at various Hartmann numbers. }
\end{figure}

A similar instability mode, which is of type IV and differs from mode
II by the opposite spanwise symmetry, appears at $\Ha\approx28.$
Although this mode becomes more unstable than mode II at $\Ha\gtrsim55,$
both modes are superseded by a more unstable mode of symmetry I, which
emerges at $\Ha\approx27$ and becomes dominating at $\Ha\gtrsim29.$
In contrast to modes II and IV, the vorticity component along the
magnetic field of mode I is an even function in the field direction.
Marginal Reynolds number and the relative phase velocity for this
mode iare shown in figure \ref{fig:rewk_Ge1_11-21}. The last instability
mode for $c=1,$ which appears at $\Ha\approx37,$ is of type III,
and it differs from mode I by the opposite spanwise symmetry. Although
the critical Reynolds number for mode III is initially significantly
higher than that for mode I, with increase of the Hartmann number
the difference between the two modes quickly diminishes and becomes
negligible at $\Ha\gtrsim100$ (see figure \ref{fig:rec_Ha_Ge1c}).
This is due to the localization of unstable perturbations at the sidewalls,
which takes place in a sufficiently strong magnetic field when a virtually
stagnant core of the flow forms. Also note that the increase of the
critical Reynolds number with the Hartmann number for modes I/III
is much slower than that for modes II/IV. This is due to the even
distribution of the vorticity component along the magnetic field which
makes modes I/III less susceptible to the magnetic field than modes
II/IV. 

For lower wall conductance ratios $c=0.1,0.01,$ instability is seen
in figure \ref{fig:rec_Ha_Ge1c} to emerge at larger Hartmann numbers.
As discussed below, this is because the finite conductivity of the
walls normal to the magnetic field becomes important at $\Ha\gtrsim c^{-1}.$

The critical Reynolds number for modes I/III can be seen in figure
\ref{fig:rec_Ha_Ge1c}(a) to increase asymptotically as $\RE_{c}\sim\Ha^{1/2}.$
This asymptotic regime requires a relatively strong magnetic field,
especially for low wall conductance ratio. Namely, for $c=0.01,$
asymptotics start to emerge at $\Ha\gtrsim10^{4}.$ The best fit including
the two subsequent terms, which are significant for $\Ha\gtrsim10^{3}$
and according to the asymptotic solution (\ref{eq:C}) are $O(\Ha^{-1/2})$
and $O(\Ha^{-1})$ relative to the leading-order term, gives $\RE_{c}\sim\kappa\Ha^{1/2}$
with $\kappa\approx109\pm1.5$ for $c=1,0.1$ and $\kappa\approx120\pm10$
for $c=0.01.$ These asymptotics imply that the relevant length scale
for the instability modes I/III is the thickness of the side layers
$\delta\sim\Ha^{-1/2}.$ This is confirmed by the critical wavenumber,
which is seen in figure \ref{fig:rec_Ha_Ge1c}(b) to increase asymptotically
as $k_{c}\sim0.5\Ha^{1/2}$ for all wall conductance ratios. Since
the frequency based on this length scale is expected to scale as $\omega\sim\Ha,$
the relative phase velocity for modes I/III, which is plotted in figure
\ref{fig:rec_Ha_Ge1c}(c), approaches a constant at a sufficiently
large Hartmann number. The best fit including the three leading terms
yields $-\omega/(\RE k)\approx0.47$ for $c=1,0.1$ and $-\omega/(\RE k)\approx0.48$
for $c=0.01.$ This means that the unstable modes I/III travel downstream
with nearly half of the maximal jet velocity. The phase velocity for
modes II/IV is seen in figure \ref{fig:rec_Ha_Ge1c}(c) to be noticeably
higher and approaches $0.8$ at $\Ha\approx100.$

\begin{figure}
\begin{centering}
\includegraphics[width=0.5\textwidth]{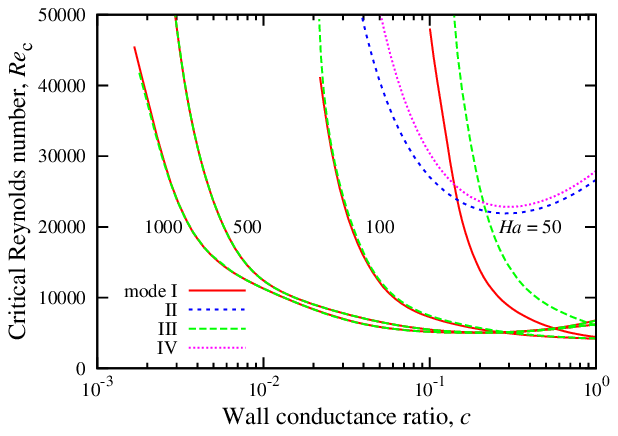}\put(-165,135){(\textit{a})}\includegraphics[width=0.5\textwidth]{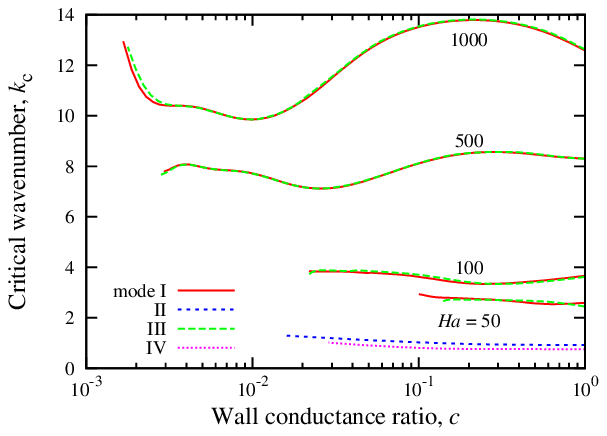}\put(-165,135){(\textit{b})} 
\par\end{centering}

\caption{\label{fig:Rec-w}The critical Reynolds number (a) and the wavenumber
(b) versus the wall conductance ratio at various Hartmann numbers. }
\end{figure}

Variation of the critical Reynolds number and the associated wavenumber
with the wall conductance ratio is shown in figure \ref{fig:Rec-w}.
As the wall conductance ratio $c$ reduces, the flow is seen to become
linearly very stable as in the duct with insulating walls. The higher
the Hartmann number, the smaller is the wall conductance ratio down
to which the instability persists. This is because the effect of $c$
is determined by the relative conductance of the Hartmann layer, which
drops directly with its thickness as $\sim\Ha^{-1}.$ Thus, for $c\ll1$
and $\Ha\gg1,$ the relevant parameter is $c\Ha$ which according
to (\ref{bc:eff}) has to be sufficiently small for the wall to be
effectively insulating. Figure \ref{fig:Rec-w}(a) indicates that
the flow becomes linearly stable when $c\Ha\lesssim1.$ However, the
stabilization of the flow does not proceed monotonically with the
reduction of the wall conductance ratio. For sufficiently large Hartmann
numbers $(\Ha\gtrsim500),$ the critical Reynolds number is seen in
figure \ref{fig:Rec-w}(a) first to drop slightly before eventually
starting to rise when the wall conductance ratio approaches $c\sim\Ha^{-1}.$
This weak destabilization of the flow is associated with the development
of the sidewall jets, which according to (\ref{eq:cmax}) attain maximum
fraction of the volume flux at $c\approx\Ha^{-1/4}$ (see figure \ref{fig:flw}b).
Thus, the lowest critical Reynolds numbers for $\Ha=500$ and $1000$
occur at $c\approx0.2$ and $0.18,$ respectively. Owing to this slight
minimum, the overall variation of $\RE_{c}$ at high Hartmann numbers
remains relatively weak down to $c\sim10^{-2}.$ 

\begin{figure}
\centering{}\includegraphics[bb=90bp 90bp 275bp 255bp,clip,width=0.4\textwidth]{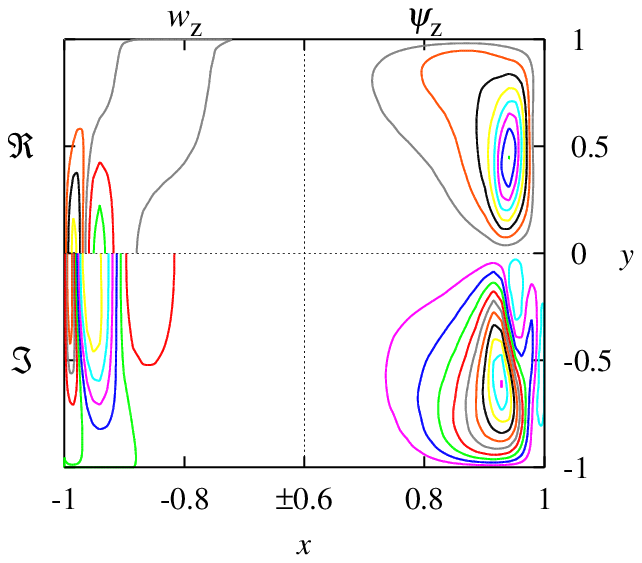}\put(-165,155){(\textit{a})}\includegraphics[bb=0bp 20bp 250bp 240bp,clip,width=0.5\textwidth]{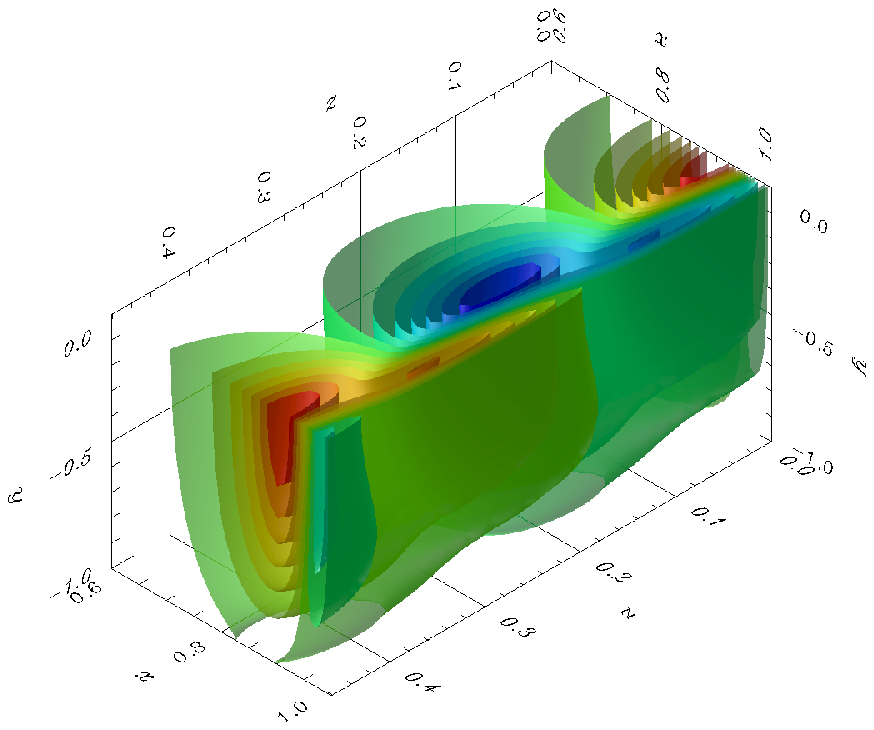}\put(-165,155){(\textit{b})}\caption{\label{fig:xy-egv}Amplitude distributions of the real $(y>0)$ and
imaginary $(y<0)$ parts of $\hat{w}$ $(x<0)$ and $\hat{\psi}_{z}$
$(x>0)$ (a) and the isosurfaces of $\hat{\psi}_{y}$ of the critical
perturbation of mode I in the vicinity of side-wall over half height
of duct cross-section for $\protect\Ha=10^{3}$ the wall conductance
ratio $c=0.1.$ }
\end{figure}

The complex amplitude distribution of the critical mode I streamwise
velocity perturbation $\hat{w}$ $(x<0)$ and the same streamfunction
component $\hat{\psi}_{z}$ $(x>0),$ which both are mirror-symmetric
with respect to the vertical mid-plane $x=0,$ are shown in figure
\ref{fig:xy-egv}(a) for $c=0.1$ and $\Ha=10^{3}.$ These two quantities
define respectively the potential and solenoidal components of the
flow perturbation in the cross-section of the duct. At this high Hartmann
number, the perturbations are seen to be localized at the sidewalls
and effectively separated by a virtually unperturbed fluid core. This
makes the perturbation pattern for mode I practically indistinguishable
from that for mode III, which differs from the former only by the
opposite $x$ symmetry.

\begin{figure}
\centering{}\includegraphics[bb=110bp 90bp 370bp 260bp,clip,width=0.5\columnwidth]{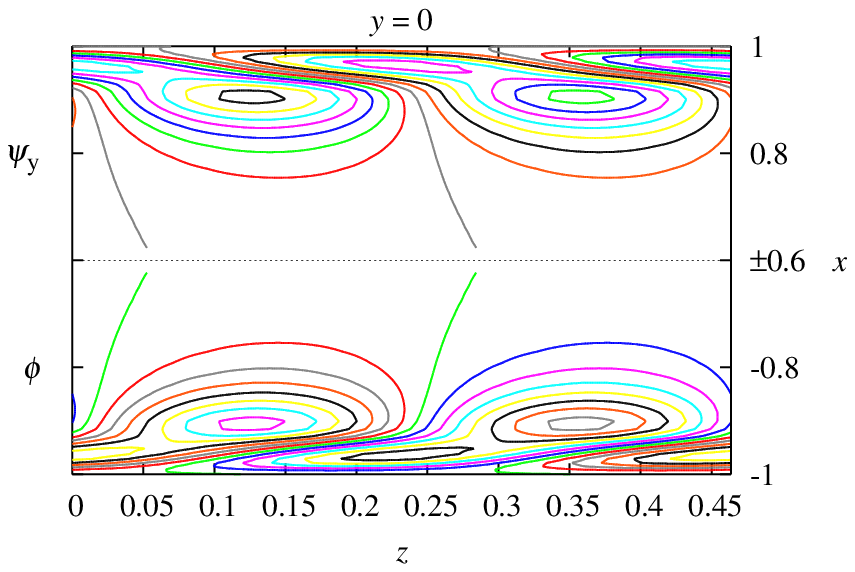}\put(-200,110){(\textit{a})}\includegraphics[bb=110bp 90bp 370bp 260bp,clip,width=0.5\columnwidth]{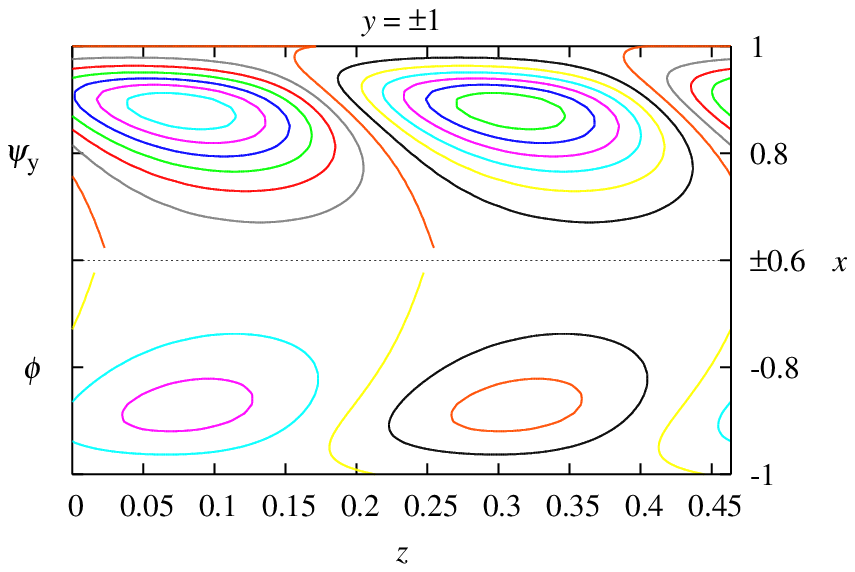}\put(-200,110){(\textit{b})}\\
\includegraphics[bb=110bp 90bp 370bp 260bp,clip,width=0.5\columnwidth]{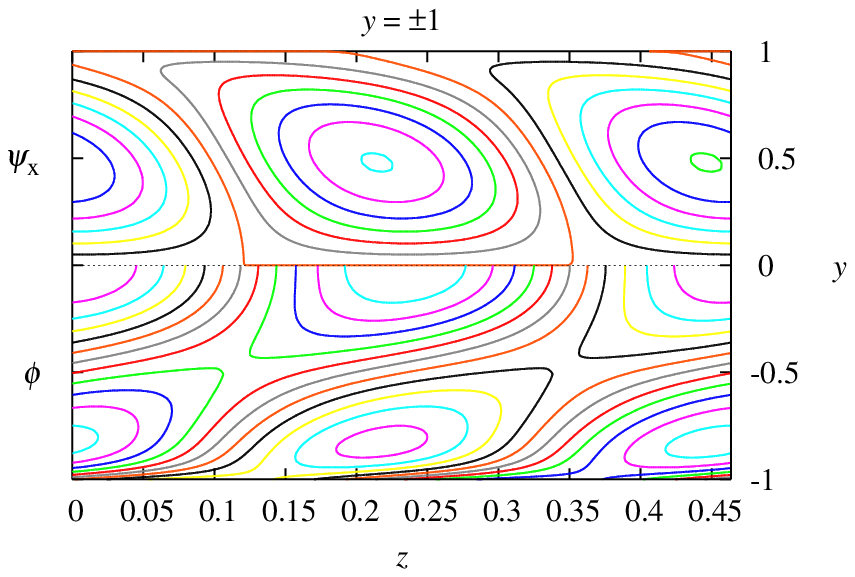}\put(-200,110){(\textit{c})}\caption{\label{fig:iso}Isolines of $\psi_{y}$ $(x>0)$ and $\phi$ $(x<0)$
in the mid-plane $(y=0)$ (a) and at the horizontal walls $(y=\pm1)$
(b); isolines of $\psi_{x}$ $(y>0)$ and $\phi$ $(y<0)$ at the
side walls $(x=\pm1)$ of the duct for the critical perturbation of
mode I at $\protect\Ha=10^{3}$ and $c=0.1.$ The isoline spacing
in (b) as well as that for $\psi_{x}$ in (c) is by a factor of $10$
smaller, whereas for $\phi$ in (c) it is by a factor $100$ smaller
than the original spacing in (a). }
\end{figure}

The real and imaginary parts, which are plotted in the upper $(y>0)$
and lower $(y<0)$ halves of the duct, are defined by the normalization
condition $\int_{S}\Re[\vec{\hat{v}]}^{2}\,\mathrm{d}s=\int_{S}\Im[\vec{\hat{v}]}^{2}\,\mathrm{d}s,$
where the integrals are taken over the duct cross-section $S.$ The
latter two quantities define perturbation distributions over the duct
cross-section shifted in time or in the streamwise direction by a
quarter of a period or wavelength, respectively. The isolines of $\hat{\psi}_{z}$
show the streamlines of the rotational flow perturbation component
in the cross-sectional plane. Circulation for mode I is seen to be
dominated by one vortex that stretches along the sidewall in each
quadrant of the duct. This vortex advects the non-uniformly distributed
momentum in the sidewall jets, so giving rise to the streamwise velocity
perturbation $\hat{w}$ which is shown on the left-hand side of figure
\ref{fig:xy-egv}(a). Since the latter varies streamwise, it gives
rise to a converging flow perturbation over tthe cross-section where
the base flow is accelerated and to a diverging one where the base
flow is slowed down. This, in turn, gives rise to a circulation in
the plane transverse to the magnetic field, i.e. in the $xz$-plane,
which is defined by $\psi_{y}.$ The spatial pattern of this quantity
is shown in figure \ref{fig:xy-egv}(b) over the wavelength for one
quadrant of the duct. The lines of constant $\psi_{y}$ correspond
to the streamlines of flow perturbation in the $xz$-plane. Note that
the spatial distribution of $\psi_{y}$ is very similar to that of
the electric potential perturbation $\phi.$ This is because both
quantities are governed by identical equations (\ref{eq:psih},\ref{eq:phih}),
and differ only by the boundary conditions (\ref{eq:bcx},\ref{eq:bcy}).
As seen in figure \ref{fig:iso}(a,b), the isolines of $\psi_{y}$
$(x>0)$ and $\phi$ $(x<0)$ are nearly identical in the horizontal
mid-plane $(y=0)$, whereas they differ more from each other at thehorizontal
walls $(y=\pm1).$ Analogous distributions of $\psi_{x}$ $(y>0)$
and $\phi$ $(y<0)$ at the sidewalls $(x=\pm1)$ of the duct can
be seen in figure \ref{fig:iso}(c). Note that the apparent flow implied
by the streamlines at the solid walls, where according to the no-slip
condition flow must be absent, are due to the streamfunction decomposition.
In contrast to the velocity components, which satisfy the boundary
conditions separately and the solenoidality condition all together,
each streamfunction component defines a solenoidal velocity field
that satisfies only the impermeability condition whereas the no-slip
condition is satisfied by all streamfunction components together.

The component of the flow perturbation associated with the vorticity
along the magnetic field, which is shown in figure \ref{fig:xy-egv}(b),
dominates the instability in a sufficiently strong magnetic field.
Namely, for -$\Ha=10^{3}$ and the wall conductance ratios $c$ from
$1$ to $0.01,$ the $y$-component of vorticity-stream function carries
$99\%$ to $97\%$ of the perturbation energy, 
\[
E\propto\int_{S}\hat{\left|\vec{v}\right|}^{2}\,\mathrm{d}s=\int_{S}\Re[\hat{\vec{\omega}}\cdot\hat{\vec{\psi}}^{*}]\,\mathrm{d}s,
\]
whereas the $y$-component of the velocity perturbation carries respectively
$1\%$ to $3\%$ of $E.$ This suggests that the flow perturbation
in a high magnetic field is dominated by a single streamfunction-vorticity
component aligned with the magnetic field.

\section{\label{sec:sum}Summary and conclusions}

We have presented analytical and numerical results concerning MHD
flow and its linear stability in a square duct with thin conducting
walls. First, we extended the original asymptotic solution obtained
by \citet{Walker1981} for the base flow in a high magnetic field
to moderate magnetic fields. The asymptotic solution, which was confirmed
by the numerical results, showed that the fraction of the volume flux
carried by the side layers attains its maximum at the wall conductance
ratio $c\sim\Ha^{-1/4}.$ For $c\ll1,$ this implies that an extremely
strong magnetic field with $\Ha\sim c^{-4}$ is required for the sidewall
jets to fully develop.

In a square duct with moderate wall conductance ratio $c=1,$ the
flow was found to become linearly unstable at $\Ha\gtrsim10.$ The
vorticity component along the magnetic field for the first instability
mode is an odd function in the field direction and an even function
in the spanwise direction. There is another slightly more stable mode
that differs from the previous one by the opposite spanwise symmetry.
The increase of the magnetic field, which smooths out the flow perturbation
along the flux lines, results in a strong damping of these two instability
modes. In a strong magnetic field, the instability is dominated by
two additional modes which differ from the previous two by the even
vorticity distribution along the magnetic field. This makes the latter
two modes more uniform along the magnetic field and thus less susceptible
to damping. The spanwise symmetric mode is slightly more unstable
than the antisymmetric mode. This is because the former consists of
the vortices at the opposite sidewalls which are co-rotating and thus
enhancing each other through a shared flux, whereas the contrary is
the case for the latter mode. This difference becomes insignificant
in a strong magnetic field, where the opposite vortices  are effectively
separated by a virtually unperturbed core flow. In this case, the
critical Reynolds number based on the jet velocity and the associated
wavenumber for both modes increase asymptotically as $\RE_{c}\sim110\Ha^{1/2}$
and $k_{c}\sim0.5\Ha^{1/2},$ while the modes travel downstream with
a phase speed close to half of maximal jet velocity. These asymptotics
imply that the instability takes place in the sidewall jets with the
characteristic thickness $\delta\sim\Ha^{-1/2}$ which determines
the actual length scale of the instability.

When the magnetic field is sufficiently strong, the reduction of the
wall conductance ratio results in a weak destabilization of the flow
until $c\sim\Ha^{-1/4},$ where the critical Reynolds number attains
a minimum and then starts to increase as $c$ is reduced further.
This slight minimum of $\RE_{c}$ corresponds to the wall conductance
ratio discussed above at which the fraction of the volume flux carried
by the sidewall jets attains a maximum. The critical Reynolds number
becomes very high at $c\lesssim\Ha^{-1},$ which corresponds to the
wall conductance ratio smaller than that of the Hartmann layer. At
this point, the walls become effectively insulating, which leads to
the transformation of the sidewall jets into the Shercliff layers
\citep{Shercliff1953} The latter are expected to become linearly
unstable like the Hartmann layer at $\RE_{c}\sim10^{4}\Ha^{1/2}$
\citep{Potherat2007}, which is too high to be reliably computed using
the current numerical method. It may also be of little significance
as the flow in the insulating duct is known to be turbulent at much
lower Reynolds numbers \citep{Shatrov2010}. 

This, however, is not necessarily the case for the flows with sidewall
jets, which have relatively low critical Reynolds numbers. Namely,
the critical Reynolds number $\RE_{c}\sim91\Ha^{1/2}$ for Hunt's
flow, which has jets very similar to those in the duct with thin walls,
is a bit lower than that found in this study. Both flows also have
very close critical wavenumbers and phase speeds. In contrast, the
flow in a perfectly conducting duct, which has much weaker jets, has
a critical Reynolds number $\RE_{c}\sim642\Ha^{1/2},$ which is significantly
higher than that for the two other flows. Finally, rescaling our critical
Reynolds number with the total volume flux (\ref{eq:Q}) in a square
duct $(A=1)$ with a low wall conductance ratio $c\ll1,$ we obtain
$\bar{\RE}_{c}\approx520.$ Although this result is a factor of $1.7$
higher than $\bar{\RE}_{c}\approx313$ found by \citep{Ting1991},
it is still significantly lower than the Reynolds number at which
turbulence is observed in experiments as well as in direct numerical
simulations of this type of flow \citep{Kinet2009}. This discrepancy,
which is opposite to what is observed in the Hartmann flow as well
as in purely hydrodynamic shear flows \citep{Hagan2014}, has no convincing
explanation yet. 

\begin{acknowledgements}
This work was supported by Helmholtz Association of German Research Centres (HGF) in the framework of the LIMTECH Alliance. The authors are indebted to the Faculty of Engineering and Computing of Coventry University for the opportunity to use its high performance computer cluster. JP would like to thank S. Molokov for pointing out a number of relevant previous publications.
\end{acknowledgements}
\appendix

\section{\label{sec:bflow}Asymptotic solution for the base flow}

The principal characteristics of the base flow are best revealed by
the asymptotic solution which is briefly reproduced below. In contrast
to \citet{Walker1981}, we use the induced magnetic field instead
of the electric potential. This allows us to obtain a more general
solution that is valid not only for small but also for moderate and
large wall conductance ratios. We also derive a more accurate result
for the flow rate in magnetic fields of a limited strength. This correction
is important because unrealistically high magnetic fields with $\Ha\gtrsim c^{-4}\gg1$
are required for attaining the asymptotic flow regimes considered
by \citet{Walker1981}.

We start with the Hartmann layers, which have relative thickness $\sim\Ha^{-1}$
and are located in the vertical magnetic field at the top and bottom
walls $(y=\pm1)$. Following the standard approach, we introduce stretched
coordinates $\tilde{y}_{\pm}=\Ha(1\mp y)$ in which equations \ref{eq:wbar}
and \ref{eq:bbar} for the leading-order terms reduce to
\begin{align}
\partial_{\tilde{y}}^{2}\bar{w}\mp\partial_{\tilde{y}}\bar{b} & =0,\label{eq:wbar1}\\
\partial_{\tilde{y}}^{2}\bar{b}\mp\partial_{\tilde{y}}\bar{v} & =0.\label{eq:bbar1}
\end{align}
These equations define the well-known exponential velocity profile
in the Hartmann layers, which complement the outer velocity distribution
$\bar{w}_{0}(x,y)$ to be determined in the following by ensuring
the no-slip conditions at $y=\pm1:$ 

\begin{equation}
\bar{w}(x,y)=\bar{w}_{0}(x,y)-\bar{w}_{0}(x,\pm1)e^{-\Ha(1\mp y)}.\label{sol:wbar}
\end{equation}
The composite solution for the magnetic field can be written similarly
as 
\begin{equation}
\bar{b}(x,y)=\bar{b}_{0}(x,y)-\bar{b}_{\pm}e^{-\Ha(1\mp y)},\label{sol:bbar}
\end{equation}
where the constants $\bar{b}_{\pm}=(1+c_{n}\Ha)^{-1}\left.(\bar{b}_{0}\pm c_{n}\partial_{y}\bar{b}_{0})\right|_{y=\pm1}$
follow from the boundary condition (\ref{bc:bbar}) with $c_{n}$
standing for the conductance ratio of the Hartmann walls. The latter
may in general be different from the conductance ratio of the parallel
walls which is denoted in the following by $c_{\tau}.$ Finally, taking
into account that the coefficients of the exponential terms in (\ref{sol:wbar})
and (\ref{sol:bbar}) are related to each other through equations
(\ref{eq:wbar1}) and (\ref{eq:bbar1}), we obtain an effective boundary
condition for the outer variables 
\begin{equation}
\left[\bar{w}_{0}+(1+c_{n}\Ha)^{-1}(c_{n}\partial_{y}\bar{b}_{0}\pm\bar{b}_{0})\right]_{y=\pm1}=0,\label{bc:eff}
\end{equation}
which holds from an insulating $(c_{n}=0)$ up to a perfectly conducting
$(c_{n}\rightarrow\infty)$ Hartmann wall. In the following, the Hartmann
walls will be assumed to be relatively well conducting, which means
$1+c_{n}\Ha\approx c_{n}\Ha\gg1$ in the equation above. 

Now the effective boundary condition (\ref{bc:eff}) can be used to
determine the solution outside Hartmann layers. Let us focus on the
parallel layers that form along the sidewalls at $x=\pm A$ and have
characteristic thickness $\sim\Ha^{-1/2}.$ These layers can conveniently
be described using stretched coordinates $\tilde{x}=\Ha^{1/2}(A\pm x).$
Then equations (\ref{eq:wbar}) and (\ref{eq:bbar}) reduce in leading-order
terms to 
\begin{align}
\partial_{\tilde{x}}^{2}\bar{w}_{0}+\partial_{y}\bar{b}_{0} & =-1,\label{eq:wbar0}\\
\partial_{\tilde{x}}^{2}\bar{b}_{0}+\partial_{y}\bar{w}_{0} & =0.\label{eq:bbar0}
\end{align}
where the pressure gradient has been normalized to simplify the solution
as $\bar{P}=-\textit{\ensuremath{\Ha}}^{-1}.$ The boundary condition
(\ref{bc:bbar}) at the side wall then takes the form
\begin{equation}
\left.\tilde{c}_{\tau}\partial_{\tilde{x}}\bar{b}_{0}-\bar{b}_{0}\right|_{\tilde{x}=0,}=0,\label{bc:bbar0}
\end{equation}
where $\tilde{c}_{\tau}=c_{\tau}\Ha^{1/2}.$ Assuming, as usual, that
the effect of viscosity represented by the first term in (\ref{eq:wbar0})
is confined to the side layer, i.e. $\left.\partial_{\tilde{x}}^{2}\bar{w}_{0}\right|_{\tilde{x}\rightarrow\infty,}\rightarrow0,$
we obtain 
\[
\left.\bar{b}_{0}\right|_{\tilde{x}\rightarrow\infty,}\rightarrow-y.
\]
This result substituted into (\ref{bc:eff}) then yields 
\begin{equation}
\left.\bar{w}_{0}\right|_{\tilde{x}\rightarrow\infty,}\rightarrow(1+c_{n}^{-1})\Ha^{-1}=\bar{w}_{\infty}.\label{eq:winf}
\end{equation}
The same condition applies along the whole Hartmann wall, i.e. $\bar{w}_{0}(\tilde{x},\pm1)=\bar{w}_{\infty},$
as long as $\bar{b}_{0}\approx-y$, which according to (\ref{eq:wbar0})
requires $\bar{w}_{0}\ll1.$ In this case, the solution can be sought
as
\begin{align}
\bar{w}_{0}(\tilde{x},y) & =\bar{w}_{\infty}[1+\sum_{k=0}^{\infty}w_{k}(\tilde{x})\cos(\kappa y)],\label{eq:w0}\\
\bar{b}_{0}(\tilde{x},y) & =-y+\bar{w}_{\infty}\sum_{k=0}^{\infty}b_{k}(\tilde{x})\sin(\kappa y),\label{eq:b0}
\end{align}
where $\kappa=\pi(k+\frac{1}{2}).$ Substituting this into (\ref{eq:wbar0})
and (\ref{eq:bbar0}), we obtain
\begin{align}
w_{k}''+\kappa b_{k} & =0,\label{eq:wk}\\
b_{k}''-\kappa w_{k} & =0.\label{eq:bk}
\end{align}
The no-slip boundary condition $\bar{w}_{0}(0,y)=0$ and the thin-wall
condition (\ref{bc:bbar0}) take the form 
\begin{align}
w_{k}(0) & =-\int_{0}^{1}\cos(\kappa y)\thinspace\mathrm{d}y/\int_{0}^{1}\cos^{2}(\kappa y)\thinspace\mathrm{d}y=(-1)^{k+1}2/\kappa,\label{bc:wk}\\
\bar{w}_{\infty}[\tilde{c}b_{k}'(0)-b_{k}(0)] & =-\int_{0}^{1}y\sin(\kappa y)\thinspace\mathrm{d}y/\int_{0}^{1}\sin^{2}(\kappa y)\thinspace\mathrm{d}y=(-1)^{k}2/\kappa^{2}.\label{bc:bk}
\end{align}
 The solution of (\ref{eq:wk}) and (\ref{eq:bk}) decaying away from
the sidewall can be written as 
\begin{align*}
w_{k}(\tilde{x}) & =e^{-\lambda\tilde{x}}B\left[C\sin(\lambda\tilde{x})+\cos(\lambda\tilde{x})\right],\\
b_{k}(\tilde{x}) & =e^{-\lambda\tilde{x}}B\left[C\cos(\lambda\tilde{x})-\sin(\lambda\tilde{x})\right],
\end{align*}
where $\lambda=\sqrt{\kappa/2},$ and $B=(-1)^{k+1}2/\kappa$ and
$C=-(\bar{w}_{\infty}^{-1}\kappa^{-1}+\tilde{c}_{\tau}\lambda)/(1+\tilde{c}_{\tau}\lambda)$
are the constants defined by the boundary conditions (\ref{bc:wk})
and (\ref{bc:bk}). For perfectly conducting walls, corresponding
to $c_{n}=c_{\tau}\rightarrow\infty$, we have  while in the case
of Hunt's flow, corresponding to $c_{n}\rightarrow\infty$ and $c_{\tau}=0$,
we have $C=-\Ha/\kappa.$ For the thin sidewall satisfying $\tilde{c}_{\tau}=c_{\tau}\Ha^{1/2}\gg1,$
we have
\begin{equation}
C\sim-\frac{\bar{w}_{\infty}^{-1}\kappa^{-1}}{\tilde{c}_{\tau}\lambda}+\frac{\bar{w}_{\infty}^{-1}\kappa^{-1}}{(\tilde{c}_{\tau}\lambda)^{2}}-1,\label{eq:C}
\end{equation}
where the second and third terms represent higher-order small corrections.
Although the last term becomes significant for relatively well conducting
walls satisfying $\tilde{c}_{\tau}/\Ha=c_{\tau}/\Ha^{1/2}\gtrsim1,$
it cancels out in the expression for the volume flux carried by the
side layer, which takes the form 
\[
q=\Ha^{-1/2}\int_{0}^{\infty}\int_{0}^{1}\left(\bar{w}_{0}(\tilde{x},y)-\bar{w}_{\infty}\right)\mathrm{d}y\thinspace\mathrm{d}\tilde{x}\sim\frac{\Ha^{-1/2}}{3\tilde{c}_{\tau}}(1-\alpha\tilde{c}_{\tau}^{-1}),
\]
where $\alpha=6(32-\sqrt{2})\pi^{-9/2}\zeta(\frac{9}{2})\approx1.1206$
and $\zeta(x)$ is the Riemann zeta function \citep{Abramowitz1964}.
Taking into account that $\tilde{c}_{\tau}q\rightarrow0$ when $\tilde{c}_{\tau}\rightarrow0,$
we can write
\[
q\sim\frac{\Ha^{-1/2}}{3\tilde{c}_{\tau}}(1+\alpha\tilde{c}_{\tau}^{-1})^{-1},
\]
which is asymptotically valid not only for large but also for small
$c_{\tau}.$ Then the fraction of the volume flux carried by the side
layer is 
\begin{equation}
\gamma=\frac{q}{q+\bar{w}_{\infty}A}=\left(1+3A(1+c_{n}^{-1})(c_{\tau}+\alpha\Ha^{-1/2})\right)^{-1}.\label{eq:gam}
\end{equation}
For a strong magnetic field satisfying $\Ha^{-1/2}\ll c_{\tau},$
we have 
\begin{equation}
\gamma=\left(1+3Ac_{\tau}(1+c_{n}^{-1})\right)^{-1}.\label{eq:ginf}
\end{equation}
For $c_{n}\sim c_{\tau}\ll1,$ this reduces further to 
\begin{equation}
\gamma=(1+3A)^{-1},\label{eq:ginfw}
\end{equation}
 which was originally obtained by \citet{Walker1981}. This asymptotic
state, however, requires an extremely strong magnetic field, which
can be estimated as follows. For a fixed $\Ha$ and $c_{n}=c_{\tau},$
(\ref{eq:gam}) attains a maximum at 
\begin{equation}
c=\alpha^{1/2}\Ha^{-1/4}.\label{eq:cmax}
\end{equation}
 Thus, for $c=10^{-2},$ $\Ha=\alpha^{2}c^{-4}\gtrsim10^{8}$ is required
to attain this $\gamma$.

\bibliographystyle{jfm}
\addcontentsline{toc}{section}{\refname}\bibliography{Ductw}

\end{document}